\documentstyle[aps,prb,epsfig]{revtex}
\begin{document}
\author{V.Ya. Demikhovskii \footnote{demi@phys.unn.ru} and
D.V. Khomitsky}
\address{Nizhny Novgorod State University \\
 Gagarin Ave. 23, Nizhny Novgorod 603950, Russia}
\title{Quantum states and optics in a {\it p}-type heterojunction with lateral
surface quantum dot or antidot superlattice subjected to perpendicular
magnetic field}

\maketitle
\begin{abstract}
The studies of quantum states and optics in a {\it p}-type heterojunction with
lateral surface quantum dot (antidot) superlattice and in the presence of
perpendicular magnetic field are performed. For the first time the Azbel'--
Hofstadter problem is solved for holes in a complicated valence band described
by the $4 \times 4$ Luttinger Hamiltonian. The set of magnetic subbands is
obtained for separate hole levels in a wide interval of magnetic field.
We found remarkable differences between hole spectra and the Hofstadter
"butterfly" for electrons. The influence of the spin-orbit interaction onto
wavefunctions and energy spectrum has been investigated. The probabilities of
optical transitions between quantum states in the valence band and donors
located in the monolayer inside the heterojunction are calculated. The set of
parameters (superlattice periods, amplitude of periodic potential, magnitude
of magnetic field, etc.) required for experimental observation of splitted
hole Landau levels is determined.
\end{abstract}

\vspace{0.5cm}
PACS number(s): \ 73.21.-b, 73.21 Cd, 78.67.-n

\vspace{0.5cm}

\section{Introduction}

The problem of quantum states of 2D Bloch electrons subjected to magnetic
field remains actual over several last decades. The fascinating physical
phenomena occurring here are caused by the mutual effects of the crystalline
periodic potential and the non-periodic vector potential of uniform magnetic
field. The former leads to the energy band structure while the latter tends
to form discrete energy levels. The crucial parameter determining the nature
of quantum states in this problem is magnetic flux $\Phi$ penetrating
the lattice elementary cell. If the flux is equal to rational number $p/q$
of flux quanta $\Phi_0=2\pi \hbar c/|e|$ where $p$ and $q$ are mutually prime
integers, it is possible to define a new set of translations on the lattice,
called magnetic translations \cite{Zak,LP} for which the quasimomentum is
a good quantum number. To be specific, let the vector potential of uniform
magnetic field $B$ be chosen in Landau gauge
\begin{equation}
\label{gauge}
{\bf A}=(0,Bx,0)
\end{equation}

\noindent and $\Phi/ \Phi_0=p/q$. Under such conditions the simplest form of
magnetic translations on a square lattice with the period $a$ is
\begin{eqnarray}
\label{mtr}
x\rightarrow x+qna, \quad  y\rightarrow y+ma
\end{eqnarray}

\noindent where $n$ and $m$ are integers. From (\ref{mtr}) it follows that
an elementary cell in the presence of magnetic field (now called a magnetic
cell) is $q$ times larger in $x$ direction, and the corresponding magnetic
Brillouin zone (MBZ) is defined as
\begin{eqnarray}
\label{MBZ}
-\pi /qa \le k_x \le \pi /qa, \quad  -\pi /a  \le k_y \le \pi /a.
\end{eqnarray}

\noindent The electron wavefunction gains an additional phase under
the magnetic translations. The relation between the translated and the initial
wavefunctions in magnetic field is known as the generalized Bloch conditions
(or Peierls conditions) \cite{Peierls,LP}
\begin{eqnarray}
\nonumber
\psi_{k_x k_y}(x,y,z)=\psi_{k_x k_y}(x+qa,y+a,z)\exp(-ik_x qa)\times \\
\label{pei}
\times \exp(-ik_y a)\exp(-2\pi ipy/a).
\end{eqnarray}

\noindent When the amplitude of periodic potential $V_0$ is smaller than
the cyclotron energy $\hbar \omega _c$ one can neglect the influence of
neighboring Landau levels and may obtain the set of magnetic subbands arising
from a single level \cite{Thouless}. Although the magnitude of magnetic field
is represented in terms of fraction $p/q$ everywhere in the discussed theory,
it should be stressed that the numerator $p$ and the denominator $q$ also
exhibit themselves separately. In particular, every Landau level splits into
$p$ subbands with degeneracy degree $q$ which means that the number of
subbands depends only on $p$ and the size of magnetic Brillouin zone
(\ref{MBZ}) is a function only of $q$. If it becomes needful to include
the interaction between Landau levels, numerical methods are usually applied
\cite{Silb,Geisel,DP}. However, all dependencies on $p$,\  $q$,
and $p/q$ which have been mentioned in this paragraph, remain the same.

During last years several significant theoretical and experimental aspects of
the discussed problem have been investigated. In particular, quantization of
Hall conductance in 2D electron gas with additional periodic potential has
been studied \cite{Thouless,PG,Geisel,Kohmoto,Usov}. One might expect that
each of magnetic subbands gives a Hall conductance equal to $e^2/ph$, but
according to Laughlin each subband must carry an integer multiple of the Hall
current carried by the entire Landau level. In more complicated models
describing Bloch electrons in magnetic field the manifestation of quantum
chaos has been discovered \cite{Eroms,Petschel,Ketzmerick}.

Recently the number of experimental studies have been performed in order
to investigate the electron quantum states in 2D heterojunctions with lateral
surface superlattice of quantum dots (antidots). Such a system is convenient
for investigation of both classical effects (commensurability of the lattice
periods and cyclotron radius, transition to chaos, etc.) and of the energy
spectrum consisting of magnetic subbands. For example, in Refs. 11,14
the oscillations of longitudinal magnetoresistance have been detected under
the conditions where classical cyclotron radius $2R_c$ envelopes an integer
number of antidots or numerous reflections from one antidot occur. The first
experimental evidences of Landau levels splitted into the set of magnetic
subbands have been obtained by longitudinal magnetoresistance studies
\cite{Schl}. Then, the measurements of Hall resistance in a subband energy
spectrum of 2D electrons have also been performed \cite{Albrecht}.
In several recent publications the magnetotransport in 2D hole gas with
lateral periodic modulation was studied \cite{Weiss02,DK02}.

Besides the magnetotransport measurements, the attempts of magnetooptical
studies of interband transitions between the conduction band and acceptor
impurities have been performed in $n$ -type heterostructures \cite{Kukushkin}.
The experiments in $p$-type heterojunctions without periodic potential have
also become possible due to the progress in technology which substantially
improved the quality of $p$ channels in GaAs/AlGaAs heterojunctions
\cite{Volkov}. Thus, almost all intriguing phenomena found for 2D electron
systems were also observed in 2D hole channels.

The specific features of hole quantum states which have caused the interest
to them may be briefly described as non-trivial effects of symmetry and
the spin-orbit interaction. It is known that in the absence of magnetic field
the electron spectrum in a symmetrical quantum well is twofold degenerate with
respect to spin. Opposite, in an asymmetrical heterojunction grown, for
example, in $z$ direction where $V(z)\ne V(-z)$ the relativistic orbital
interaction of the electron magnetic moment and macroscopic heterojunction
potential leads to the breakdown of spin degeneracy. Only twofold Kramers
degeneracy
$E\left({\bf k},\uparrow \right)=E\left({\bf -k},\downarrow \right)$ remains.
In order to obtain transparent and valuable results from transport and optical
experiments, one may need to choose the set of parameters (superlattice
periods, value of magnetic field and amplitude of periodic potential, etc.)
which provide a sharp, easily  distinguishable picture of non-overlapped
magnetic subbands originating from a particular Landau level. Such energy
spectra and wavefunctions are studied in the present paper together with
calculation of luminescence intensities for transitions between magnetic
subbands and impurities. In Sec. II we study the hole quantum states in a
$p$ - type heterojunction subjected to magnetic field only (Subsec. II A)
and both to magnetic field and the periodic potential of quantum dot
superlattice (Subsec. II B). The spin-orbit coupling is included here which is
principal for the description of holes in semiconductors. For the first
time the calculation of magnetic subbands and four-component wavefunctions
for holes in a heterojunction is performed. It was found that the structure
of hole magnetic subbands differs from the well-known Hofstadter "butterfly",
especially at high magnetic field. Then in Sec. III we calculate the matrix
elements and luminescence intensities for direct optical transitions between
hole magnetic subbands and electrons bound to donors which are located in
the monolayer inside the heterojunction. The huge difference in the magnitude
of luminescence intensities corresponding to different magnetic subbands was
found and the dependence of transition probabilities on polarization was
investigated. These results may be used for identification of complicated
magnetic subband spectra in magnetooptical experiments. The summary of our
results is given in Sec. IV.

\section{Hole quantum states in the presence of lateral superlattice and
magnetic field}

\subsection{Hole Landau levels in a $p$ -type heterojunction without
periodic potential}

Let us now consider the upper edge of GaAs $p$-like valence band near
the $\Gamma_8$-point ${\bf k}=0$. In the presence of the external magnetic
field ${\bf B}$ in the $\langle 001 \rangle$ direction (hereafter denoted by
$z$), the effective Hamiltonian $H_L$ (neglecting linear $k$ terms) is
obtained from the $4\times 4$ Luttinger Hamiltonian \cite{Luttinger,Broido} by
replacing the components of the wave vector by their operator forms,
\begin{equation}
\label{kmag}
k_{\alpha}\rightarrow \hat{k_{\alpha}}=-i\frac{\partial}{\partial x_{\alpha}}
+\frac{e}{c}A_{\alpha}.
\end{equation}

\noindent Besides, one has to include the $\kappa$ terms, which represent
the interaction of the electron's spin magnetic moment with the external
magnetic field. In this section the atomic units $\hbar=m_0=1$ are used.
Writing $H_L$ in terms of the creation and destruction operators,
\begin{equation}
\label{aa}
a^+=\frac{R_c}{\sqrt{2}}k_+, \qquad a=\frac{R_c}{\sqrt{2}}k_-
\end{equation}

\noindent where $k_{\pm}=k_x \pm ik_y$, $R_c=\left[\frac{c}{eB}\right]^{1/2}$,
and making the no-warping approximation, one obtains the $H_L$ in
the following form:
\begin{eqnarray}
\label{lutt}
\vspace{1 cm}
H_L= \left[ \matrix{
H_{11} & {\overline \gamma}\sqrt{3}(eB/c)a^2 & \gamma_3\sqrt{6eB/c}\ k_z a
& 0 \cr
\ & H_{22} & 0 & -\gamma_3\sqrt{6eB/c}\ k_z a \cr
\ & \ & H_{33} & {\overline \gamma}\sqrt{3}(eB/c)a^2 \cr
\ & \ & \ & H_{44}
} \right],
\end{eqnarray}

\noindent where

\begin{eqnarray}
\nonumber
H_{11}=-(\gamma_1/2-\gamma_2)k_z^2-(eB/c)\left[(\gamma_1+\gamma_2)
\left(a^+ a + \frac{1}{2} \right)+\frac{3}{2}\kappa \right], \\
\nonumber
H_{22}=-(\gamma_1/2+\gamma_2)k_z^2-(eB/c)\left[(\gamma_1-\gamma_2)
\left(a^+ a + \frac{1}{2} \right)-\frac{1}{2}\kappa \right], \\
\nonumber
H_{33}=-(\gamma_1/2+\gamma_2)k_z^2-(eB/c)\left[(\gamma_1-\gamma_2)
\left(a^+ a + \frac{1}{2} \right)+\frac{1}{2}\kappa \right], \\
\nonumber
H_{44}=-(\gamma_1/2-\gamma_2)k_z^2-(eB/c)\left[(\gamma_1+\gamma_2)
\left(a^+ a + \frac{1}{2} \right)-\frac{3}{2}\kappa \right].
\end{eqnarray}

\noindent The lower half of matrix (\ref{lutt}) is obtained by
Hermitian conjugation. The hole energy here is measured as negative,
$e$ is a modulus of elementary charge,
${\overline \gamma}=(\gamma_2+\gamma_3)/2$. The band parameters
appearing in (\ref{lutt}) are taken from Ref.22:
\ $\gamma_1=6.85$, $\gamma_2=2.1$, $\gamma_3=2.9$, and  $\kappa=1.2$.
The Luttinger Hamiltonian (\ref{lutt}) is written in a basis of $p$-like
atomic functions $v_j({\bf r})$ which transform as a set of eigenfunctions
of the angular momentum operator $J=3/2$. These $|J;m_J\rangle$ basis
functions may be written as following:
\vspace{0.5cm}
\begin{equation}
\label{jmj}
\cases{
v_1=\mid\frac32;\frac32\rangle=\left|-\sqrt{1/2}(x+iy)\uparrow
\right\rangle, &  \cr
v_2=\mid\frac32;-\frac12\rangle=\left|-\sqrt{1/6}(x-iy)\uparrow
-\sqrt{2/3}z\downarrow\right\rangle, &  \cr
v_3=\mid\frac32;\frac12\rangle=\left|\sqrt{1/6}(x+iy)\downarrow
-\sqrt{2/3}z\uparrow\right\rangle, & \cr
v_4=\mid\frac32;-\frac32\rangle=\left|-\sqrt{1/2}(x-iy)\downarrow
\right\rangle, & \cr }
\end{equation}
\vspace{0.5cm}

\noindent where the arrows indicate the $z$-projection of spin.
The holes in GaAs/AlGaAs $p$-type heterojunction grown in $z$ direction which
is parallel to the magnetic field are confined by the smoothly varying
potential $V_h(z)$ which allows us to apply the effective-mass
approximation. The potential $V_h(z)$ is of a triangular shape, and on
the boundary $\psi(0)=0$. It should be noted that such a shape does not have
the inversion symmetry, i.e. $V_h(z)\ne V_h(-z)$ which leads to the breakdown
of the twofold spin degeneracy and to the splitting of energy levels of
the effective-mass Hamiltonian
\begin{equation}
\label{heff}
H_{eff}=H_L(a^+,a,k_z)+V_h(z) \cdot {\hat E}
\end{equation}

\noindent even at the absence of magnetic field \cite{Broido}. Hereafter
${\hat E}$ stands for a unit $4\times 4$ matrix. The lack of
inversion symmetry of the atomic potential of GaAs crystal lattice is present
also in bulk material and is described by linear $k$-terms in Luttinger
Hamiltonian. However, the effects caused by these terms (the displacement of
subband maximum in {\bf k}-space \cite{Rashba,Sham}) are negligible compared
with those induced by heterostructure potential and thus are not considered
here.

We first observe that for $B=0$ the Hamiltonian (\ref{heff}) becomes diagonal
with elements
\begin{eqnarray}
\nonumber
H_h=-(\gamma_1/2-\gamma_2)\frac{d^2}{dz^2}+V_h(z), \\
\nonumber
H_l=-(\gamma_1/2+\gamma_2)\frac{d^2}{dz^2}+V_h(z)
\end{eqnarray}

\noindent that yields an infinite set of doubly degenerate heavy and light
hole subband energies and eigenfunctions $c_{\nu_j}(z),\nu =1,2,\ldots$.
These functions are usually obtained by solving Schr\"odinger and Poisson
equations self-consistently. As a result, the shape of potential $V(z)$ has
a varying gradient which reflects the changes in electric field inside
the heterojunction \cite{Volkov,Broido}. Thus, the precise shape of functions
$c_{\nu_j}(z)$ differs from the one for the case of uniform electric field.
However, the investigations of energy spectrum and the matrix elements of
transitions between 2D Bloch quantum states and impurities require only
the information about the overlapping between different localized functions
$c_{\nu_j}(z)$, and between them and well-known wavefunctions of impurities.
The intervals of localization for $c_{\nu_j}(z)$ can be obtained with high
accuracy for all subbands of size quantization considered in this paper since
the shape of $V(z)$ in a single GaAs/AlGaAs heterojunction is well-known
\cite{Broido}. The solution of the effective-mass equation with
the Hamiltonian (\ref{heff}) may be written as a four-component vector of
envelope functions in the $|J;m_J\rangle$ basis (\ref{jmj}). As it was shown
by Luttinger \cite{Luttinger}, in the presence of magnetic field and under
axial approximation one can distinguish the eigenstates of operator
(\ref{heff}) by a discrete quantum number $N$ which defines the particular
set of Landau quantum states. These states have $k_y$-component of momentum
under Landau gauge (\ref{gauge}). In the presence of the heterostructure
potential the $k_z$-component in (\ref{heff}) is replaced by the operator
$k_z=-i\partial / \partial z$. Hence, an eigenstate $F_{Nk_y}$ of
the operator (\ref{heff}) consists of four envelope functions
and the hole wavefunction is written as
\begin{equation}
\nonumber
\Psi_{Nk_y}=\sum_{j=1}^4 F_{jNk_y}v_j
\end{equation}

\noindent where $v_j$ is a $|J;m_J\rangle$ basis function and
\begin{equation}
\label{fn}
F_{Nk_y}=e^{ik_y}\left(C_1(z)u_{N-2}, \ C_2(z)u_{N},\ C_3(z)u_{N-1},
\ C_4(z)u_{N+1} \right).
\end{equation}

\noindent In Eq.(\ref{fn}) $u_N(x)$ is a harmonic oscillator wavefunction
and the envelope functions $C_j(z)$ are constructed as a superposition of
the zero-field functions $c_{\nu_j}(z)$ with numerically defined coefficients,
which vanish for negative indexes $N$. For example, for $N=-1$ one can
obtain $F_{-1}=(0,0,0,C_4(z)u_0)$, for $N=0$ the solution
$F_0=(0,C_2(z)u_0,0,C_4(z)u_1)$, and for $N \ge 2$ all four components of
(\ref{fn}) will be nonzero \cite{choice}. After substituting the function
(\ref{fn}) into the Schr\"odinger equation with the Hamiltonian (\ref{heff})
one obtains an algebraic eigenvalue problem. We restrict ourselves to
the first three levels of size quantization which corresponds to two
heavy- and one light-hole levels. This approximation seems to be valid in
heterojunctions with typical hole concentration $n_h=5 \times 10^{11} cm^{-2}$
and depletion-layer density $N_{dep}=10^{15} cm^{-3}$
for which only the lowest hole level is occupied \cite{Broido,Volkov}. For
each level of size quantization we take into account several Landau levels
shown on Fig.1. Here one can see the electron-like behavior of light-hole
Landau levels at low magnetic field caused by proximity of the second
heavy-hole subband. We assume that the introduction of periodic potential
with the amplitude $V_0$ (see the following Subsec.) does not change
$c_{\nu_j}(z)$ significantly since $|V_0|$ considered in our paper is much
smaller than the size quantization energies. Hence, in our further studies we
use the matrix elements of the effective Hamiltonian (\ref{heff}) calculated
for the functions $c_{\nu_j}(z)$ and the size quantization energies which
have been discussed above.
\begin{figure}[t]
\leavevmode
\centering{\epsfbox{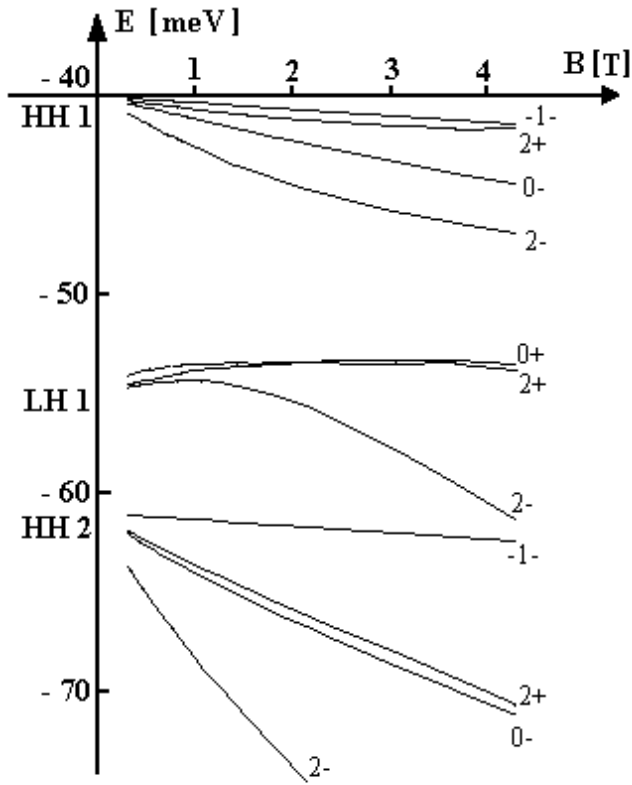}}
\caption{Set of hole Landau levels corresponding to the first three subbands
of size quantization (two heavy-hole levels HH1, HH2, and one light-hole level
LH1). The electron-like behavior of the light-hole Landau levels at low
magnetic field can be observed. Each level is characterized by Landau index
$N=-1,0,1,\ldots$ and by dominating spin projection $\pm$ (see text).
Hereafter the energy is measured from the top of the valence band in bulk
GaAs.}
\end{figure}

\subsection{Bloch quantum states in the presence of lateral surface
superlattice}

The problem of hole quantum states in a $p$-type heterojunction subjected
to magnetic field and affected by a lateral superlattice is described by
the Schr\"odinger equation with the vector potential (\ref{gauge}) and the
2D periodic potential of a lateral superlattice which can be chosen in
the form \cite{DP}
\begin{equation}
\label{vxy}
V(x,y)=V_0\cos^2\frac{\pi x}{a}\cos^2\frac{\pi y}{a}.
\end{equation}

\noindent Here $a$ is the superlattice period and the case $V_0<0\ (>0)$
corresponds to the periodic electric potential generated by quantum dot
(antidot) superlattice. The Hamiltonian for magnetic Bloch hole quantum
states is a sum of (\ref{heff}) and (\ref{vxy}):
\begin{equation}
\label{htot}
H=H_{eff}+V(x,y)\cdot {\hat E},
\end{equation}

\noindent The eigenvectors of the operator (\ref{htot}) are four-component
envelope functions written in the $|J;m_J\rangle$ basis (\ref{jmj}):
\begin{equation}
\label{psienv}
\Psi^{envelope}_{k_x k_y}({\bf r})=\left(\psi^{(1)}_{k_x k_y}({\bf r}),\
\psi^{(2)}_{k_x k_y}({\bf r}),\ \psi^{(3)}_{k_x k_y}({\bf r}),\
\psi^{(4)}_{k_x k_y}({\bf r})  \right),
\end{equation}
\noindent and the total hole wavefunction is
\begin{eqnarray}
\nonumber
\Psi_{k_x,k_y}({\bf r})=
\psi^{(1)}_{k_x k_y}({\bf r})\left|\frac32;\frac32\right\rangle+
\psi^{(2)}_{k_x k_y}({\bf r})\left|\frac32;-\frac12\right\rangle+ \\
\label{psihole}
\psi^{(3)}_{k_x k_y}({\bf r})\left|\frac32;\frac12\right\rangle+
\psi^{(4)}_{k_x k_y}({\bf r})\left|\frac32;-\frac32\right\rangle.
\end{eqnarray}

\noindent The crucial statement here is the following: as long as
the periodic potential (\ref{vxy}) is applied, every hole envelope function
in Eq.(\ref{psienv}) becomes a magnetic Bloch function classified by $k_x$
and $k_y$ quantum numbers varying in the MBZ (\ref{MBZ}). It should be
mentioned that the translation properties of each component of the envelope
function (\ref{psienv}) in $(xy)$ plane are the same as for the single-
component electron wavefunction. In particular, every component of
(\ref{psienv}) satisfies to the Peierls condition (\ref{pei}). Hence, one can
write the components of (\ref{psienv}) as a superposition of the Landau
quantum states as it was demonstrated in Refs.4,8, namely
\begin{eqnarray}
\nonumber
\psi^{(j)}_{k_x k_y}({\bf r})=\frac{1}{La\sqrt{q}}
\sum_{\nu_j}c_{\nu_j}(z)\sum_{N_j}\sum_{n=1}^p G_{j \nu_j N_j n}(k_x,k_y)
\sum_{l=-L/2}^{L/2}u_{Nj}\left(\frac{x-x_0-lqa-nqa/p}{\ell_H} \right)\times
\\
\label{psiho}
\times \exp\left(ik_x\left[lqa+\frac{nqa}{p}\right]\right)
\exp\left(2\pi iy\frac{lp+n}{a}\right)\exp(ik_y y),
\end{eqnarray}

\noindent where for a particular $|J;m_J\rangle$ projection $j$ we take
into account several levels of size quantization $\nu_j$ and for each of them
we assume several Landau levels $N_j$. Then, analogous to the electron problem
described, for example, in Refs 5-7,9, after substituting the wavefunction
(\ref{psihole}) into the Schr\"odinger equation with the Hamiltonian
(\ref{htot}) one obtains the eigenvalue problem for the coefficients
$G_{j \nu_j N_j n}(k_x,k_y)$ and the hole magnetic subbands
$\varepsilon_{\nu_j N_j n}(k_x,k_y)$:
\begin{equation}
\label{matrix}
\sum_{j'\nu_j'N_j'n'}\left(
H^{j'\nu_j'N_j'n'}_{j \nu_j N_j n}+V_{j \nu_j N_j n}^{j'\nu_j'N_j'n'}
(p/q,k_x,k_y) \right)G_{j'\nu_j'N_j'n'}=\varepsilon G_{j \nu_j N_j n}.
\end{equation}

\noindent Here the notation $H^{j'\nu_j'N_j'n'}_{j \nu_j N_j n}$ is used for
the projection of the Hamiltonian (\ref{heff}) onto our basis
$(j\ \nu_j\ N_j\ n)$ and $V_{j \nu_j N_j n}^{j'\nu_j'N_j'n'}(p/q,k_x,k_y)$
stands for the matrix elements of the periodic potential (\ref{vxy})
calculated in this basis. We have diagonalized the system (\ref{matrix}) for
different values of magnetic field and different amplitudes of periodic
potential. The maximum size of Hermitian matrix in (\ref{matrix}) was up to
$220\times 220$. The corresponding energy spectra and hole wavefunctions
are shown on Figures 2 -- 6.

It was mentioned previously that hole Landau levels may be classified into
groups of the effective Hamiltonian (\ref{htot}) eigenvalues labeled by
the common index $N=-1,0,1, \ldots$. For example, for $N=0$ such a group
belonging to the subband of size quantization with $\nu =1$ consists of one
heavy- and one light-hole level. These levels can be obtained by
the diagonalization of $2\times 2$ matrix and are labeled by $N=0-(+)$
(see Fig.1). When the periodic potential of lateral superlattice is introduced,
the $2\times 2$ matrix yields $2p\times 2p$ matrix equation (\ref{matrix})
which spectrum consists of $2p$ magnetic subbands originating from $N=0-(+)$
levels. If the amplitude $|V_0|$ is small enough to neglect the influence of
other levels neighboring with the levels $N=0-(+)$, it is possible to study
their splitting separately. The set of $2p$ magnetic subbands originating from
the levels $N=0-(+)$ splitted by the periodic potential with $V_0=-3 meV$ is
shown on Fig.2a(b) versus the reciprocal magnetic flux $q/p$. Comparing Fig.2
and the ordinary Hofstadter "butterfly" for 2D electrons, one can see that at
low magnetic field $q/p\approx 1$ the hole spectrum looks similar to
the electron one, in particular, the clustering of hole magnetic subbands is
the same. At high magnetic fields $q/p \ll 1$ one can see the down and up
shifts of the energy on Fig.2 with respect to the center of the "butterfly" at
$q/p\approx 1$. This difference between hole and electron spectra is caused by
the off-diagonal elements of the Luttinger Hamiltonian (\ref{lutt}) which
become more significant at high magnetic fields.
\begin{figure}[t]
\leavevmode
\centering{\epsfbox{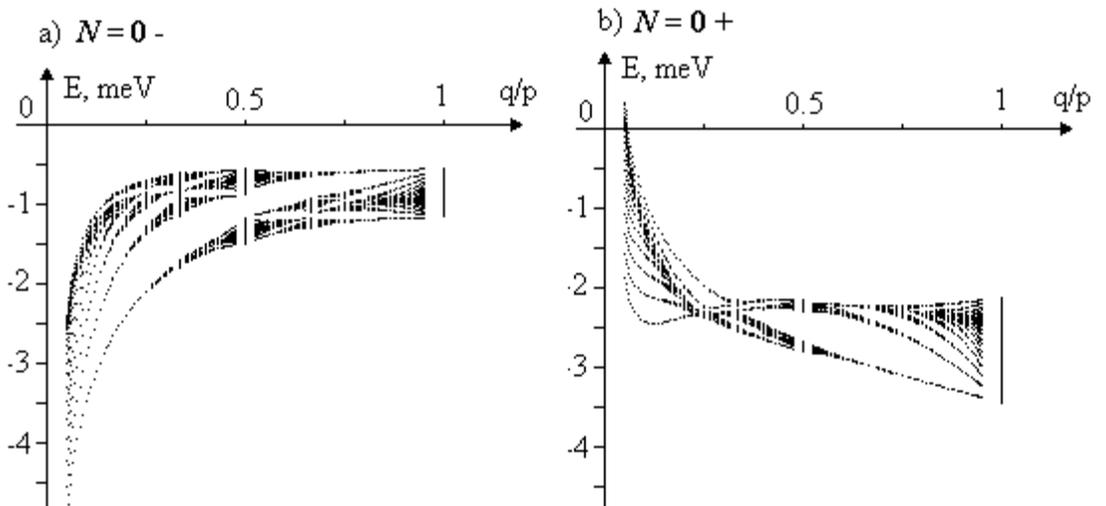}}
\caption{Energy spectrum of 2D hole gas laterally modulated by the quantum
dot superlattice  with $a=80 nm$ plotted vs reciprocal magnetic flux $q/p$.
The spectrum is shown for two hole levels $N=0-(+)$ coupled by
the off-diagonal term of the Luttinger Hamiltonian and splitted by
periodic potential (\ref{vxy}) with the amplitude $V_0=-3 meV$.}
\end{figure}

Hereafter we are going to be interested mainly in the hole energy
spectra at high (and fixed) magnetic field $p/q=20$ which corresponds to
$B\approx 12$ T. The spectrum of system (\ref{matrix}) for such magnetic
flux and for $k_x=k_y=0$ is shown on the bottom part of Fig.3 for the case of
non-overlapped subbands related to the highest hole levels $N=2+$ and $N=-1-$.
Here the sign $+(-)$ refers to the spin projection of the dominating component
of $|J;m_J\rangle$  basis \cite{Broido,Volkov}. Similar to the electron
spectra \cite{PG,DP}, every hole Landau level has splitted into $p$ narrow
magnetic subbands (which look like discrete levels) grouped near
the unperturbed Landau level (marked as a dark circle on Fig.3). Note that
the $N=2+$ and $N=-1-$ levels on Fig.3 have exchanged their positions in
energy with respect to Fig.1 which is due to the level crossing occurred
at some intermediate $p/q$. The condition $|V_0|\le\Delta E_{12}$ where
$\Delta E_{12}$ is the distance between the levels $N=2+$ and $N=-1-$ allows
to observe the set of non-overlapped magnetic subbands for these levels at
high magnetic fields.
\begin{figure}[t]
\leavevmode
\centering{\epsfbox{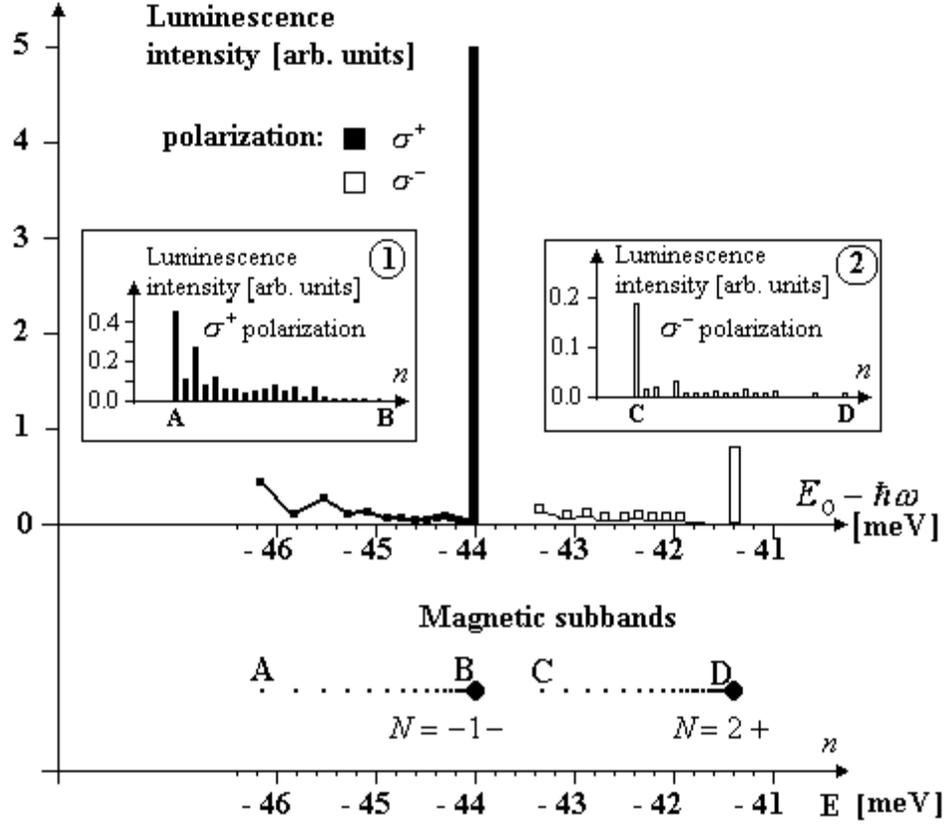}}
\caption{(Bottom) non-overlapped hole magnetic subbands related to the highest
Landau levels $N=2+$ and $N=-1-$ (black circles) at $p/q=20$ splitted by
the periodic potential of quantum dot superlattice with the amplitude
$V_0=-2.5 meV$. (Top) intensities of the luminescence with $\sigma^+$ (black)
and $\sigma^-$ (white) polarization for transitions between donor levels and
hole subbands plotted vs. the photon energy $\hbar \omega$ together with
the intensities of transitions to the unperturbed ($V_0=0$) hole levels.
The latter are shown as single bars above the level positions. The insets 1
and 2 show the fine structure of the transition intensities with respect to
subband number, and the donor energy $E_0$ is counted from the $\Gamma_8$
point of GaAs valence band.}
\end{figure}
\clearpage

In the following Sec. we will calculate the matrix elements for transitions
between the valence band and donors located in the heterojunction and
thus the knowledge of hole wavefunction in a superlattice cell is required.
On Fig.4a,b we plot the $Re \psi_j>0$ (top) and the $|\psi_j|^2$ (bottom) for
the hole wavefunction component $m_J=-3/2$ at $k_x=k_y=0$ for subbands {\it A}
and {\it B} which are indicated on Fig.3. We've not shown the contour plots
for the imaginary part of the wavefunction since they demonstrate
qualitatively the same behavior. It should be noted that for subbands
located between {\it A} and {\it B} (including themselves) the other
$|J;m_J\rangle$ components in (\ref{psihole}) are negligible under the
conditions of a non-overlap with other subbands. This is a consequence of
a single-component structure of the eigenvector $F_{-1}=(0,0,0,C_4(z)u_0)$
corresponding to the $N=-1-$ Landau level shown on Fig.1. The impact of other
components in our model can be provided only by the external periodic
potential $V(x,y)$ leading to the Landau level coupling, but this coupling is
negligible for the non-overlapped subbands shown of Fig.3, and thus no other
$|J;m_J\rangle$ components are present on Fig.4. As for the single-component
electron quantum states \cite{DP}, $Re \psi_j$ (and $Im \psi_j$ also) have
different structure with respect to the subband number $n$. Namely, in the
subband {\it A} located far from the clustering point, the $Re \psi_j$ (upper
part of Fig.4a) has much less oscillations than the $Re \psi_j$ for the
subband {\it B} belonging to the clustering region (upper part of Fig. 4b).
It should be noted that the probability density distribution (bottom parts
of Fig.4a,b) is always smooth and square symmetric, despite the possible
oscillatory character of $Re \psi_j$ or $Im \psi$. One can see that
the probability density shown on bottom parts of Fig.4a,b always has the $C_4$
symmetry which corresponds to the symmetry of the superlattice, while
$Re \psi_j$ can have lower symmetry (upper part of Fig.4b) because of the
non-symmetrical Landau gauge (\ref{gauge}). It can be expected that the
discussed difference in the $Re \psi_j$ shape in different subbands should be
reflected in the magnitude of matrix elements for transitions from donors and
it will be proved in the following Sec.
\begin{figure}[t]
\leavevmode
\centering{\epsfbox{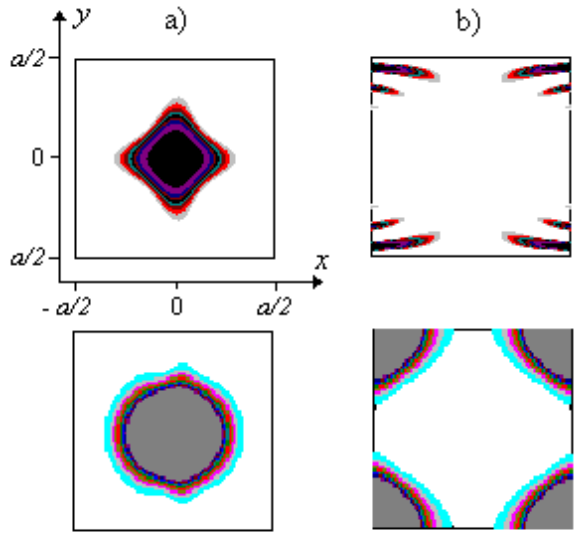}}
\caption{Envelope hole wavefunctions $Re \psi_j>0$ (top) and probability
density distributions (bottom) for the component $m_J=-3/2$ plotted in one
superlattice cell at $k_x=k_y=0$ for subbands {\it A} (Fig.4a) and {\it B}
(Fig.4b) shown on Fig.3.}
\end{figure}

When the condition $|V_0|<\Delta E_{12}$ is not fulfilled, the structure of
hole spectrum looks different. The spectrum for $V_0=-10 meV$ and
$\Delta E_{12}\approx 2.5 meV$ is shown on the bottom part of Fig.5.
\begin{figure}[t]
\leavevmode
\centering{\epsfbox{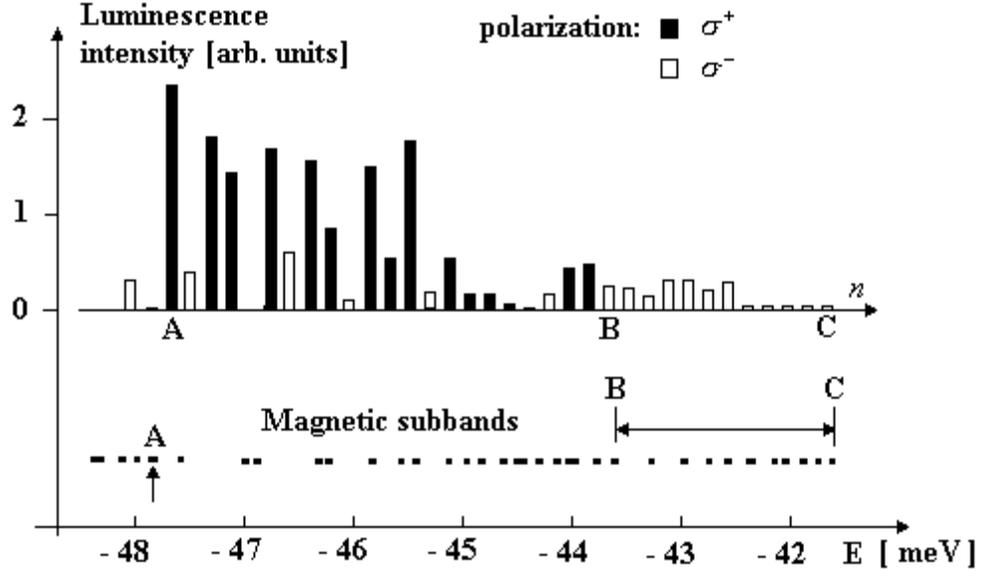}}
\caption{(Bottom) overlapped hole magnetic subbands related to the same Landau
levels but splitted by the periodic potential with larger amplitude
$V_0=-10 meV$. (Top) intensities of the luminescence with $\sigma^+$ (black)
and $\sigma^-$ (white) polarization for transitions between the monolayer of
donors and hole subbands plotted vs the subband number $n$.}
\end{figure}

In this case the magnetic subbands originating from different hole Landau
levels are strongly overlapped almost everywhere except the region near the
highest Landau level. This region is marked on Fig.5 and it contains magnetic
subbands from {\it B} to {\it C} belonging to the Landau level $N=2+$.
In this interval of non-overlapping subbands one may expect a distinguishable
behavior of transition intensities for these subbands (see the following
Sec). Under the conditions of strong subbands overlap the domination of one
of $|J;m_J\rangle$ basis component becomes less pronounced.
This is illustrated on Fig.6 where we plot the probability distributions
$\mid \psi_j \mid ^2$ for all four $|J;m_J\rangle$ components of
the wavefunction (\ref{psihole}) in the subband {\it A} marked by an arrow
on Fig.5. It is clearly seen that all the components have the same order which
is a consequence of the overlapping between the magnetic subbands originating
from Landau levels with different dominating wavefunction components.
As for the case of non-overlapped subbands which has been discussed above,
the probability density distributions shown on Fig.6 have the $C_4$ symmetry
for all $|J;m_J\rangle$ components.
\begin{figure}[t]
\leavevmode
\centering{\epsfbox{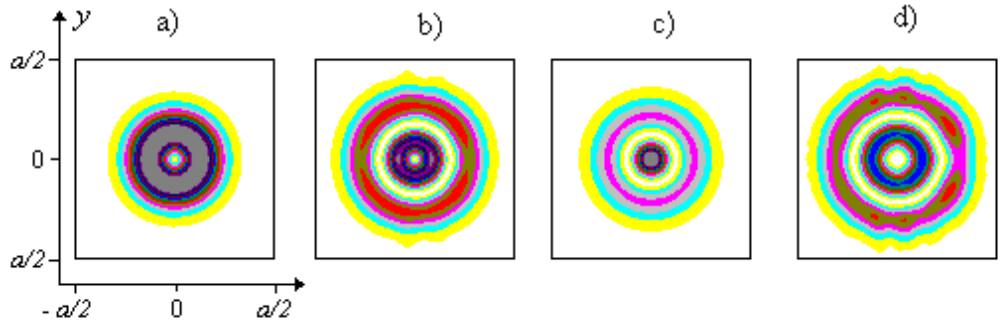}}
\caption{Probability distribution $\mid \psi _j \mid ^2$ of four hole
wavefunction components at $k_x=k_y=0$ for the subband {\it A} marked by
an arrow on Fig.5. The figures (a) -- (d) correspond to the components
$\mid\frac32;\frac32\rangle$, $\mid\frac32;-\frac12\rangle$,
$\mid\frac32;\frac12\rangle$ and $\mid\frac32;-\frac32\rangle$ of
$|J;m_J\rangle$ basis, respectively.}
\end{figure}
\clearpage

\section{Luminescence intensities for donors - valence band transitions}

As it was mentioned in the Introduction, one of the experimental methods
for investigation of quantum states in magnetic subbands is a magnetooptical
measurement of transition intensities. Below we calculate the matrix elements
and intensities of transitions between the electrons bound to the monolayer of
shallow donors and the magnetic Bloch hole states in the valence band.

Let us consider a process in which photon is emitted and the electron is
dropped from the donor atom to the valence band. Following Ref.20,
we suppose that the monolayer of donors is located at a well-defined distance
from the heterojunction interface. The final quantum state $\Psi^f_{k_xk_y}$
is the hole wavefunction (\ref{psihole}), and the initial quantum state
$\Psi^i$ is a hydrogen-like wavefunction of a donor impurity located inside
the heterojunction at the distance $z=z_0$ from the interface and at the
point $(x_0,y_0)$ in the superlattice cell. This impurity is described by
the envelope function $\psi_D({\bf r},{\bf r}_0)$ where
${\bf r}_0=(x_0,y_0,z_0)$ and
\begin{eqnarray}
\nonumber
\psi_D({\bf r},{\bf r}_0)=
A\exp\left(-\frac{1}{r_D}\left[(x-x_0)^2+(y-y_0)^2+(z-z_0)^2\right]^{1/2}
\right),
\end{eqnarray}

\noindent where $r_D=\kappa_e \hbar^2/m^*e^2$ stands for the donor Bohr radius
in GaAs/AlGaAs heterostructure with the dielectric constant $\kappa_e$ and
the effective mass at the bottom of conduction band $m^*$. The value of $r_D$
obtained from luminescence measurements is about 15 nm \cite{Volkov}.
The conduction band is characterized by an $s$-type atomic function
$s_{\alpha}({\bf r})$ where the index $\alpha=1(2)$ corresponds to the
function $\mid s\uparrow\rangle \left(\mid s\downarrow\rangle \right)$.
Since the total ensemble of donor atoms does not have a definite projection of
an angular momentum, one can write
\begin{eqnarray}
\nonumber
\Psi^i=\psi_D({\bf r},{\bf r}_0)\cdot
\frac{\mid s\uparrow\rangle+\mid s\downarrow\rangle}{\sqrt2}.
\end{eqnarray}

\noindent After the definition of initial and final quantum states one can
write the magnetoluminescence intensity $I(\hbar \omega)$ as
\begin{eqnarray}
\label{intens}
I(\hbar \omega) \propto {\sum_{if}{\overline {\mid M_{if} \mid ^2}}
\delta (E_f-E_i-\hbar \omega) }
\end{eqnarray}

\noindent where we assume that the initial state $\Psi^i$ is fully occupied
and the final state $\Psi^f_{k_xk_y}$ is empty. If the energy of electrons
bound to donors is fixed, the summation over the initial states in
(\ref{intens}) reduces to the multiplication by the total number of donor
atoms. The matrix element for transition between a donor and a hole state
is \cite{Ancilotto}
\begin{eqnarray}
\nonumber
M_{if}=\langle\Psi^f_{k_xk_y} \mid {\bf p\cdot e}\mid \Psi^i
\rangle= \\
\label{holemel}
=\sum_{\alpha=1}^2\sum_{j=1}^4\langle v_j \mid {\bf p\cdot e}
\mid s_{\alpha} \rangle \langle \psi^{(j)}_{k_x k_y} \mid \psi_D\rangle+
\sum_{\alpha=1}^2\sum_{j=1}^4 \langle\psi^{(j)}_{k_x k_y}
\mid{\bf p \cdot e}\mid \psi_D\rangle \langle v_j \mid s_{\alpha} \rangle,
\end{eqnarray}

\noindent where ${\bf e}$ being a unit vector in the direction of
electric field and the scalar products are defined as
\begin{eqnarray}
\nonumber
\langle v_j \mid (\ldots) \mid s_{\alpha} \rangle=
\int_{cell}v_j({\bf r})^* (\ldots) s_{\alpha}({\bf r})d{\bf r}, \\
\nonumber
\langle \psi^{(j)}_{k_x k_y}\mid(\ldots)\mid \psi_D\rangle=\int_{crystal}
\psi^{(j)*}_{k_x k_y}({\bf r})(\ldots)\psi_D({\bf r})d{\bf r}.
\end{eqnarray}

\noindent The first term in (\ref{holemel}) corresponds to the matrix elements
of the transitions from donors to the valence band. The second term vanishes
due to the orthogonality of the $|J;m_J\rangle$ functions $v_j({\bf r})$ and
$s_{\alpha}({\bf r})$ being $p$- \ and $s$-type functions, respectively.
It should be stressed that the transitions from donors to hole subbands
belonging to different Landau levels are characterized by different
polarization. On the one hand, it is a consequence of different contribution
of the $|J;m_J\rangle$ basis components into the hole quantum state
(\ref{psihole}) and, on the other hand, it is due to the fact that
the transitions from heavy holes are three times more intensive then those
from light holes (see, for example, Refs. 20,26). The overlapping of hole and
donor wavefunctions and thus the matrix element strongly depend on
the position of the donor atom in a current superlattice cell. In order to
obtain the transition probability for a superlattice with many cells we have
to average it over many possible donor positions in the $(xy)$ plane, i.e.
\begin{equation}
\label{aver}
{\overline {\mid M \mid ^2}}=\frac{1}{N_D} \sum_{x_0y_0}
\mid M(x_0y_0)\mid ^2,
\end{equation}

\noindent where $N_D$ is the total number of donor positions. We found that
due to the random position of a donor atom the matrix elements practically
(with an accuracy of few percents) do not depend on the quasimomentum which
classify the Bloch quantum state. This independence on $k_x$ and $k_y$ is
stipulated by the fact that the radius of the donor wavefunction is
considerably smaller than the superlattice period $a$. By taking this into
account, the summation over the final states in (\ref{intens}) is also
performed simply. The magnetic subbands are very narrow at $p/q=20$
and their widths are apparently smaller than the collision broadening
(the corresponding estimations for the electron gas can be found, for example,
in Ref.6). So, one can treat the magnetic subbands at $p/q=20$ as a set of
levels with fixed energy and thus also replace the summation over the final
states in (\ref{intens}) by multiplication by the total number of states in
a magnetic subband. This number is determined by the area of the magnetic
Brillouin zone (\ref{MBZ}) which is equal for all subbands. As a result, we
found that the variations of intensity (\ref{intens}) with respect to
the photon energy almost precisely repeat the behavior of matrix elements
(\ref{aver}).

On the upper side of Fig.3 we show the results for the luminescence
intensities with $\sigma^+$ and $\sigma^-$ circular polarization calculated
for two highest Landau levels $N=-1-$ and $N=2+$ at $p/q=20$ being splitted
by the periodic potential (\ref{vxy}) with the amplitude $V_0=-2.5 meV$.
On the horizontal axis we plot the photon energy $\hbar \omega$ counted from
the energy $E_0$ which is the distance between a donor level and
the $\Gamma_8$ point of the valence band. In order to compare these values
with the intensities of transitions to the unperturbed Landau levels ($V_0=0$)
we plot these intensities on the right side of each histogram of Fig.3
directly above the position of the hole Landau levels. Qualitatively,
the maximum intensity of transitions to one of $p$ magnetic subbands is
approximately $p$ times smaller than for the non-splitted Landau level which
corresponds to the ratio (equal to $p$) of the number of quantum states in one
Landau level and in one magnetic subband. It is evident that the magnetic
subbands related to different Landau levels provide the luminescence with
different polarization, just like the unperturbed hole Landau levels
\cite{Volkov}. Namely, for $\sigma^+$ polarization only the transitions to
magnetic subbands originating from $N=-1-$ level can be observed while for
$\sigma^-$ polarization the transitions to subbands from $N=2+$ level are
significant. On the insets {\it 1,2} on Fig.3 we show in detail the dependence
of the transition intensities on a subband number. One can see that
the transitions to subbands located far from the clustering point (subbands
{\it A} and {\it C}) are more intensive than for subbands near the unperturbed
Landau levels (subbands {\it B} and {\it D}). Such a behavior is
a consequence of the oscillatory character of hole wavefunctions in
{\it B}- and {\it D}-type subbands (see Fig.4). Namely, the space scale of
the wavefunction in subband {\it A} (Fig.4a) is of the same order as
the lattice constant $a$ (which is larger than the donor Bohr radius $r_D$)
while the period of wavefunction oscillations on Fig.4b is considerably
smaller than $r_D$. As a result, the matrix element for transitions in
{\it B}-type subbands decreases rapidly which explains the sharp saturation
(observable on the insets {\it 1} and {\it 2} on Fig.3) of transitions to
{\it B}- and {\it D}-type subbands compared with {\it A}- and {\it C}-type.
We believe that the described differences in magnetooptical parameters will
provide more transparence in experimental studies of hole magnetic subbands.

The behavior of transition intensities changes drastically when $|V_0|$ is
increased. As it was mentioned previously, the overlapping of magnetic
subbands occur when $|V_0|>\Delta E_{12}$. The transition intensities for
such a case are shown on the upper part of Fig.5 for the same Landau levels
$N=-1-$ and $N=2+$ at $p/q=20$ splitted by the periodic potential of quantum
dots (\ref{vxy}) with higher amplitude $V_0=-10 meV$. One can see that
the switching of polarization from $\sigma^+$ to $\sigma^-$ leads to the
total decrease of transition intensities but their dependance on
$n$ changes significantly mainly for subbands {\it B} -- {\it C} which are
not overlapped with those related to other Landau levels (see the marked
region on the bottom part of Fig.5). The switching of polarization illuminates
these subbands and thus makes possible to detect them experimentally.

\section{Summary and conclusions}

We investigated quantum states and magnetooptics of 2D holes in a {\it p}-type
heterojunction subjected to perpendicular magnetic field and periodic
potential of surface superlattice. The holes were described by the $4\times 4$
Luttinger Hamiltonian where both confinement potential and potential of
lateral surface superlattice have been introduced. This model allowed us to
figure out the influence of the spin-orbit coupling onto four-component
magnetic Bloch quantum states. We've calculated hole magnetic subbands at high
magnetic fields under consideration of several Landau levels originating from
the first three subbands of size quantization. In a wider interval of both low
and high magnetic fields the set of hole magnetic subbands originating from
two coupled Landau levels has been obtained. We found the increasing
differences with electron quantum states at high magnetic fields which are
caused by the $B$-dependent off-diagonal terms in the Luttinger Hamiltonian.
Then the calculations of matrix elements for transitions between donors and
hole magnetic subbands have been performed. We observed the characteristic
dependencies of transition intensities on a subband number and the strong
dependence on the polarization of luminescence radiation. In particular, at
$\sigma^+$ ($\sigma^-$) polarization the most intensive transitions are to
those hole magnetic subbands where "spin"-down(up) components of the
wavefunction dominate. The discussed effects allowed us to define the set of
parameters (superlattice periods, amplitude of periodic potential and magnetic
field) for possible experimental observation of sharp non-overlapping
magnetic subbands for 2D holes.

In the following paper we plan to study the magnetotransport properties of
laterally modulated 2D hole gas, and, in particular, the quantization of Hall
conductance in hole magnetic subbands.

\section*{Acknowledgments}

We thank D. Weiss, R.R. Gerhardts and D. Pfannkuche for useful discussions and
M. Kohmoto for sending us the offprint of his paper (Ref.9). This work was
supported by the RFBR (Grants No. 01-02-17102, 02-02-06440), by the Russian
Ministry of Education (Grants No. E00-3.1-413, UR 0101.020) and by the BRHE
Program (Project REC - 001).


\begin{thebibliography}{99}
\bibitem{choice}
It should be noted that the particular classification of solutions $F_N$ may
be chosen in a different way \cite{Volkov} which leads to changes in notation
only.
\end{thebibliography}
\end{document}